\documentclass[aps,twocolumn,floatfix]{revtex4}
\usepackage{graphicx}
\usepackage{amsmath}
\usepackage{color}
\newcommand{\bs}{\begin{split}}
\newcommand{\be}{\begin{equation}}
\newcommand{\ee}{\end{equation}}
\newcommand{\bea}{\begin{eqnarray}}
\newcommand{\eea}{\end{eqnarray}}
\newcommand{\bse}{\begin{subequations}}
\newcommand{\ese}{\end{subequations}}
\newcommand{\gng}{${\rm GdNiGe_3}$}
\newcommand{\gcs}{${\rm GdCu_2Si_2}$}

\newcommand{\lco}{${\rm LiCrO_2}$}
\newcommand{\cco}{$\alpha$-${\rm CaCr_2O_4}$}

\begin{document}
\title{Molecular-field-theory fits to magnetic susceptibilities of antiferromagnetic \gcs, ${\rm CuO}$, \lco, and \cco\ single crystals below their N\'eel temperatures}
\author {David C. Johnston} 
\affiliation {Ames Laboratory and Department of Physics and Astronomy, Iowa State University, Ames, Iowa 50011, USA}

\date{\today}

\begin{abstract}

A recently-developed molecular field theory (MFT) has been used to fit single-crystal magnetic susceptibility $\chi$ versus temperature~$T$ data below the respective  antiferromagnetic ordering temperature $T_{\rm N}$ for a variety of collinear and coplanar noncollinear Heisenberg antiferromagnets.  The spins in the system are assumed to interact by Heisenberg exchange and to be identical and crystallographically equivalent.  The fitting parameters for $\chi(T)$ of collinear antiferromagnets are measurable quantities: the Weiss temperature $\theta_{\rm p}$ in the Curie-Weiss law, $T_{\rm N}$, $\chi(T_{\rm N})$, and the spin~$S$\@. For coplanar noncollinear helix and cycloid structures, an additional fitting parameter is the turn angle between layers of ferromagnetically-aligned spins.  Here MFT fits to anisotropic $\chi(T)$ data from the literature for single crystals of the collinear antiferromagnets \gcs\ and CuO and the noncollinear antiferromagnets  \lco\ with a 120$^\circ$ cycloidal structure and \cco\ with a 120$^\circ$ helical structure below their respective N\'eel temperatures are presented.  The MFT fit to the anisotropic $\chi(T\leq T_{\rm N})$ data for CuO is poor, whereas the fits to the data for \gcs, \lco, and \cco\ are quite good.  The poor fit for CuO is attributed to the influence of strong quantum fluctuations associated with the small Cu spin and the quasi-one-dimensional magnetism that are not taken into account by the MFT\@.  The magnetic contribution to the zero-field heat capacity of the collinear antiferromagnet \gng\ at $T\leq T_{\rm N}$ is also fitted by the MFT.\\

\end{abstract}

\maketitle


\clearpage

\section{\label{Sec:Intro} Introduction}

A molecular-field theory (MFT) for antiferromagnetic (AF) Heisenberg spin systems containing identical crystallographically-equivalent spins has been recently formulated~\cite{Johnston2012, Johnston2015}.  A unified description of the anisotropic magnetic susceptibility $\chi(T\leq T_{\rm N})$ versus temperature $T$ below the antiferromagnetic (AF) ordering temperature $T_{\rm N}$ of both collinear and coplanar noncollinear antiferromagnets with Heisenberg exchange interactions was obtained.  The MFT is applicable to a wide range of antiferromagnets with interactions including geometric and bond-frustrating interactions that can produce a large range of the ratio
\bea
f \equiv \theta_{\rm p}/T_{\rm N}\quad ( f < 1).
\label{Eq:flimits}
\eea
Here $\theta_{\rm p}$ is the Weiss temperature in the Curie-Weiss-law fit of \mbox{$\chi(T\geq T_{\rm N})$} data in the paramagnetic (PM) regime.  This formulation of MFT uses the angles $\phi_{ji}$ between a central thermal-average moment~$i$ and those of its neighbors~$j$ with which it interacts to calculate the thermodynamic properties of antiferromagnets both above and below $T_{\rm N}$\@.  This MFT allows both collinear and coplanar noncollinear antiferromagnets to be treated on the same footing.

The second useful feature of the MFT is that it is formulated in terms of quantities that are usually easily measured or inferred with good accuracy.  For collinear antiferromagnets, these properties are $T_{\rm N}$,  $\theta_{\rm p}$, the above ratio~$f$, and the spin $S$ of the local magnetic moment.  For coplanar noncollinear helical or cycloidal antiferromagnets an additional parameter is the wave vector {\bf k} directed along the helix or cycloid axis which can be determined independently using neutron-diffraction measurements or left as a parameter that can be ob tained by fitting the anisotropic $\chi(T\leq T_{\rm N})$ data by the MFT\@.  For compounds containing other coplanar AF structures, the MFT can be used to fit single-crystal $\chi(T)$ data if the AF structure and an exchange interaction model are specified as was done for ${\rm GdB_4}$ in~\cite{Johnston2012}.  For specific exchange-interaction models, the exchange interactions $J_{ij}$ between spins~$i$ and~$j$ can be derived from the measured values of $T_{\rm N}$, $\chi(T_{\rm N})$, $\chi(T=0)$, and $\theta_{\rm p}$, which can complement information obtained from magnetic inelastic-neutron-scattering measurements.  

Subsequent papers discussed the influences of various anisotropies on the predictions of the MFT, including magnetic-dipole anisotropy~\cite{Johnston2016}, anisotropy arising from a classical anisotropy field~$H_{\rm A}$~\cite{Johnston2017b}, and quantum-mechanical uniaxial anisotropy~\cite{Johnston2017}. The $T=0$ phase diagrams in the $H_x$-$H_{\rm A}$ plane for helices with different turn angles in magnetic fields~$H_x$  applied transverse to the $z$-axis helix wave vector  with both infinite~\cite{Johnston2017} and finite~\cite{Johnston2019, Johnston2020} classical XY anisotropy fields were also obtained.  Some of these results were utilized to fit high-field magnetization data for single crystals of the helical antiferromagnet EuCo$_2$P$_2$~\cite{Johnston2017, Johnston2019} and the collinear antiferromagnet ${\rm CaCo_{1.86}As_2}$~\cite{Anand2014b}.

Modeling of experimental $\chi(T\leq T_{\rm N})$ data for coplanar noncollinear single-crystal antiferromagnets satisfying the assumptions of the MFT can help to identify and quantify when quantum fluctuations due to a small spin quantum number, frustration effects, and/or a low spin-lattice dimensionality are especially important to the physics.  Such quantum fluctuations beyond MFT can cause significant deviations of the observed $\chi(T\leq T_{\rm N})$ from the predictions of MFT, and hence such deviations can be used as a diagnostic for the importance of quantum fluctuations in a particular material.

In Ref.~\cite{Johnston2012}, in addition to the coplanar noncollinear antiferromagnet GdB$_4$ noted above,  the anisotropy in \mbox{$\chi(T\leq T_{\rm N})$} of the collinear antiferromagnets GdNiGe$_3$ and MnF$_2$ and of the coplanar noncollinear triangular 120$^\circ$ antiferromagnets YMnO$_3$ and RbCuCl$_3$ were fitted within the unified MFT\@.  Comparisons were also carried out of the powder-averaged MFT predictions with measured $\chi(T <  T_{\rm N})$ data for polycrystalline samples with inferred collinear and noncollinear AF structures~\cite{Samal2013, Goetsch2013, Samal2014, Anand2014, Nath2014, Goetsch2014}.  More recently, the anisotropic susceptibilities below $T_{\rm N}$ of single crystals of the helical antiferromagnets ${\rm EuCo_2P_2}$~\cite{Sangeetha2016}, ${\rm EuCo_2As_2}$~\cite{Sangeetha2018}, and ${\rm EuNi_2As_2}$~\cite{Sangeetha2019} were  successfully modeled by the MFT\@.  In the present paper, we compare the theoretical MFT predictions of $\chi(T\leq T_{\rm N})$ and in one case the zero-field magnetic heat capacity $C_{\rm mag}(T)$ with experimental data from the literature for single-crystal compounds for both collinear and coplanar-noncollinear AF structures.

Novel aspects of the paper include a fit of the $ab$-plane $\chi(T\leq T_{\rm N})$ of tetragonal \gcs\ with a large spin $S = 7/2$ and a collinear AF structure with the moments aligned in the $ab$ plane, but where orthogonal AF domains are present.  The fit to the anisotropic $\chi(T\leq T_{\rm N})$ of CuO shows a very large deviation of the fit from the data for the field along the ordered moment direction along the quasi-one-dimensional Cu spin-1/2 chains, illustrating a strong influence of quantum spin fluctuations associated with the small spin and low dimensionality.  On the other hand, the $\chi(T\leq T_{\rm N})$ for the triangular-lattice antiferromagnets \lco\ and \mbox{\cco}\ are nearly isotropic and independent of temperature even though the Cr spin-3/2 is still rather small, as expected~\cite{Johnston2012}.

The paper is organized as follows.  For collinear AFs the $\chi$ parallel to the ordering axis is denoted as $\chi_\parallel$ and that perpendicular to it as $\chi_\perp$.  Fits of experimental $\chi_\parallel(T\leq T_{\rm N})$ data for the collinear antiferromagnet GdCu$_2$Si$_2$ with $S=7/2$ and for CuO with $S=1/2$ by the MFT predictions are presented in Sec.~\ref{Sec:ChiCollFits}. In that section  the MFT prediction that coplanar 120$^\circ$ helical or cycloidal magnetic structures have an isotropic and nearly temperature-independent $\chi(T)$ below $T_{\rm N}$ is shown to be satisfied by experimental data for crystals of the $S=3/2$ compounds LiCrO$_2$ with a 120$^{\circ}$ cycloidal AF structure and \mbox{$\alpha$-CaCr$_2$O$_4$} with a 120$^{\circ}$ helical AF structure.  For CuO with $S=1/2$, the quality of the fit is poor, suggesting a strong influence of quantum fluctuations associated with the small spin and quasi-one-dimensionality which are not taken into account by the MFT\@. Finally, a MFT fit of the magnetic contribution $C_{\rm mag}(T)$ to the heat capacity of GdNiGe$_3$ is presented.   Concluding remarks are given in Sec.~\ref{Sec:Summary}.

\section{\label{Sec:ChiCollFits} Fits of Experimental Data for Single Crystals of Collinear Antiferromagnets}

Within MFT, the magnetic susceptibility perpendicular to the ordering axis or plane of a collinear or coplanar antiferromagnet, respectively, is independent of $T$ below $T_{\rm N}$ with the value $\chi(T_{\rm N})$, so there is no need to fit $\chi_\perp(T\leq T_{\rm N})$ data.

In the following, the experimental $\chi_\parallel(T)$ data for \mbox{$T\leq T_{\rm N}$} are fitted by~\cite{Johnston2012, Johnston2015}
\be
\chi_\parallel(T\leq T_{\rm N}) = \left[\frac{1-f}{\tau^\ast(t)-f}\right]\chi(T_{\rm N}),
\label{Eq:chiparFit}
\ee
where
\be
\bs
t &= \frac{T}{T_{\rm N}},\qquad f=\frac{\theta_{\rm p}}{T_{\rm N}}, \qquad \tau^\ast(t) = \frac{(S+1)t}{3B_S^\prime(y_0)}, \\*
y_0 &=\frac{3\bar{\mu}_0}{(S+1)t}, \qquad \bar{\mu}_0 = \frac{\mu_0}{gS\mu_{\rm B}} = B_S(y_0),
\end{split}
\label{Eq:reiterate}
\ee
$B_S(y_0)$ is the Brillouin function, $B_S^\prime(y_0)\equiv [dB_S(y)/dy]_{y=y_0}$, and $\mu_0$ is the $T$-dependent ordered moment below~$T_{\rm N}$ in applied field $H=0$ that is calculated within MFT\@.

As noted above, the constraints on the types of spin lattices the MFT can address are that the spins must be identical, crystallographically equivalent, and interact by Heisenberg exchange.  The experimental input parameters are $S$, $f$, $T_{\rm N}$ and $\chi(T_{\rm N})$.  If all four values are known, as is often the case, then there are no adjustable parameters in the fit.  The values of $\bar{\mu}_0$ and $y_0$ in Eqs.~(\ref{Eq:reiterate}) are calculated for given values of $t$ and $S$ by numerically solving the last of Eqs.~(\ref{Eq:reiterate}). Then the $y_0$ value is used to calculate $B_S^\prime(y_0)$, which is inserted into the above expression for $\tau^\ast(t)$ which is then inserted into Eq.~(\ref{Eq:chiparFit}).  The calculation is repeated for as many values of $t$ as desired.

\subsection{\label{Sec:GdCu2Si2} ${\rm\bf GdCu_2Si_2}$: a Spin-7/2 Collinear Antiferromagnet with  Orthogonal Antiferromagnetic Domains}

\begin{figure}
\includegraphics [width=1.5in]{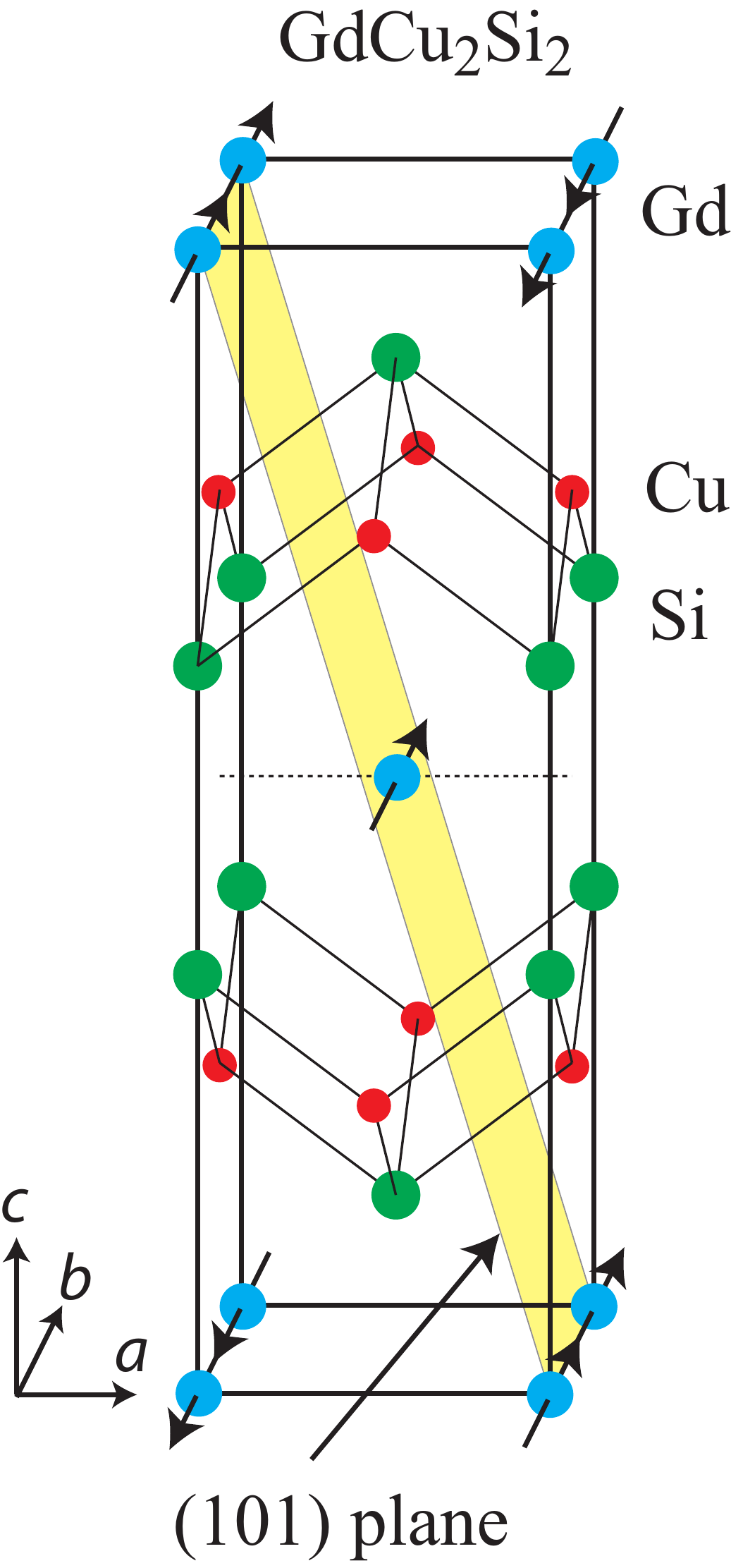}
\caption{ Crystal and magnetic structures of body-centered-tetragonal ${\rm GdCu_2Si_2}$ with the ${\rm ThCr_2Si_2}$ structure.  One crystallographic unit cell is shown.  The magnetic unit cell has dimensions $2a\times b\times 2c$ and contains four crystallographic unit cells.  The collinear magnetic structure has an AF propagation vector $(\frac{1}{2},0,\frac{1}{2})$ perpendicular to the (101) plane shown, with the magnetic moments oriented along the $b$~axis.  The Gd ordered moments are ferromagnetically aligned within each such plane.  After Ref.~\cite{Barandiaran1988}.}
\label{Fig:GdCu2Si2_Mag_struct}
\end{figure}

\begin{figure}
\includegraphics [width=3.in]{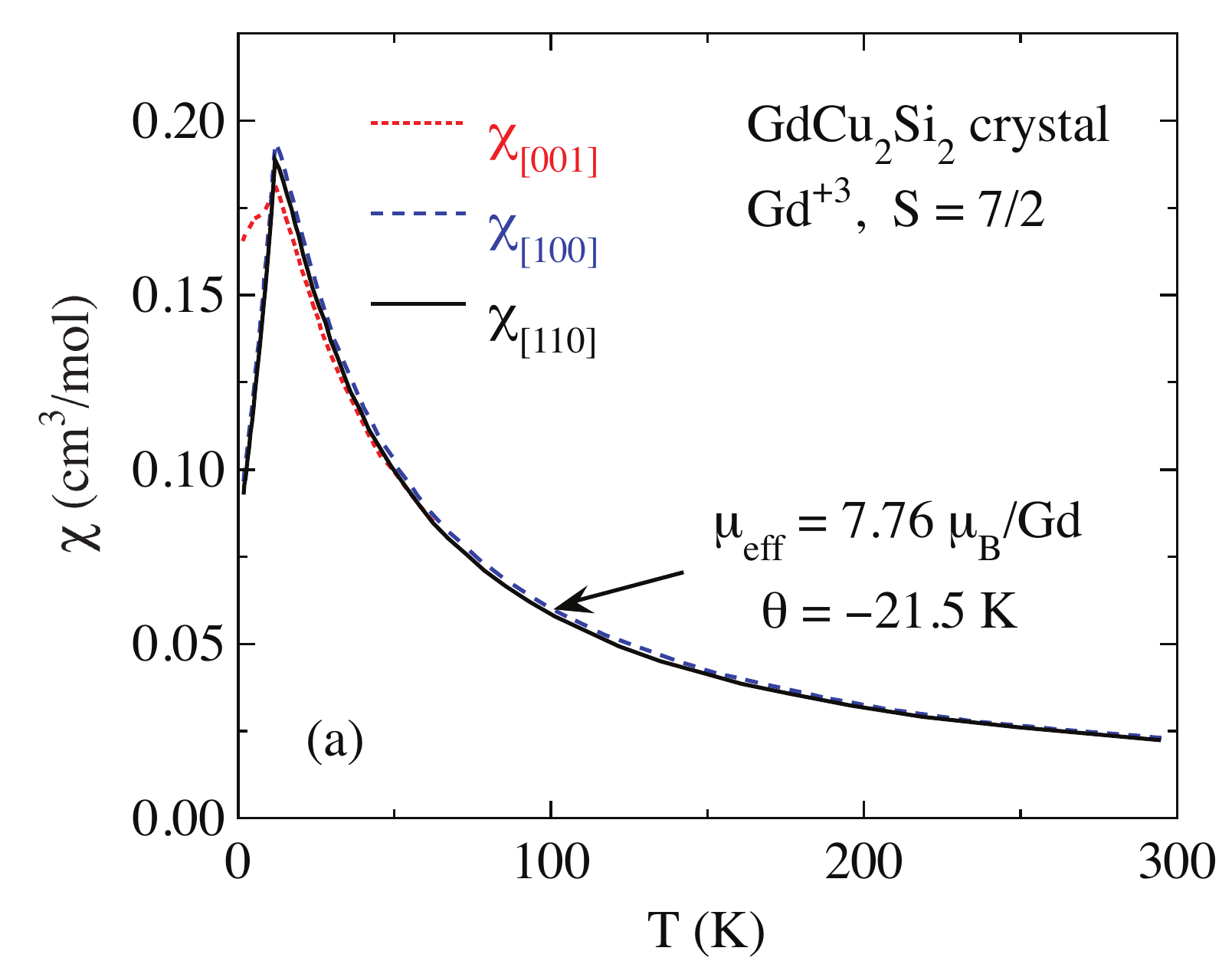}
\includegraphics [width=3.in]{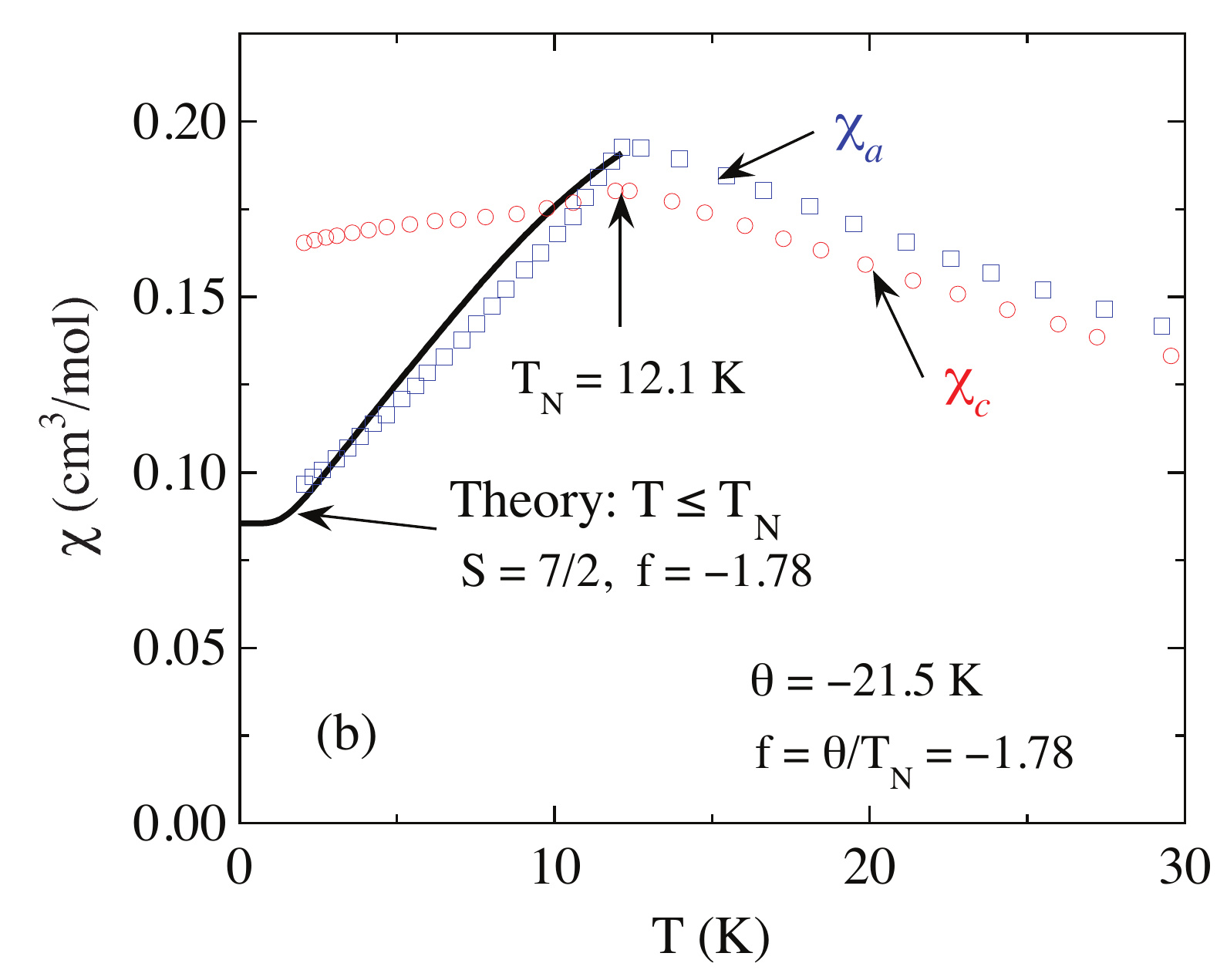}
\caption{ (a) Experimental magnetic susceptibilities $\chi$ versus temperature $T$ for a single crystal of ${\rm GdCu_2Si_2}$ with the bct ${\rm ThCr_2Si_2}$ structure~\cite{Dung2009}.  A fit to the high-temperature data by a Curie-Weiss law (not shown) gives the effective moment $\mu_{\rm eff}$ and Weiss temperature $\theta$ values  listed.  (b) Expanded plots of the $\chi(T)$ data in~(a) at low temperatures (red and blue open symbols), showing the N\'eel temperature $T_{\rm N} = 12.1$~K\@.  Also shown is the prediction of MFT in Eq.~(\ref{Eq:GdCu2Si2Fit}) for \mbox{$\chi_{a}(T\leq T_{\rm N})$} for spin $S = 7/2$ and $f = -1.78$ for equal populations of orthogonal AF domains in the $ab$~plane (solid black curve).}
\label{Fig:Dung_GdCu2Si2_chi}
\end{figure}

The compound ${\rm GdCu_2Si_2}$ has the body-centered tetragonal (bct) ${\rm ThCr_2Si_2}$-type crystal structure with space group $I4/mmm$ where the Gd atoms occupy the crystallographically-equivalent origin and body-center positions in the unit cell as shown in Fig.~\ref{Fig:GdCu2Si2_Mag_struct}.  The lattice parameters and $z$-axis Si positions were variously reported as $a=4.003$, $c=9.947$~\AA~\cite{Rieger1969} and $a=4.003$, $c=9.959$~\AA, $z_{\rm Si}=0.382$~\cite{Dung2009} at room temperature; and $a=3.922$, $c=9.993$~\AA, $z_{\rm Si}=0.368$ at 24~K~\cite{Barandiaran1988}.

The magnetic structure of ${\rm GdCu_2Si_2}$ is collinear, with the Gd ordered magnetic moments oriented along the tetragonal $b$~axis and an AF propagation vector ${\bf k}_1=(\frac{1}{2},0,\frac{1}{2})$~r.l.u.~\cite{Barandiaran1988} as shown in Fig.~\ref{Fig:GdCu2Si2_Mag_struct}.  The ordered moment at 2~K is 7.2(4)~$\mu_{\rm B}$/Gd~\cite{Barandiaran1988} in agreement with the value of $7\,\mu_{\rm B}$/Gd obtained from the relation $\mu_{\rm sat} = gS\mu_{\rm B}$, where $S=7/2$ and $g=2$.  Thus the Gd moments in (101) planes are ferromagnetically aligned and are oriented perpendicular to ${\bf k}$.  Due to the tetragonal symmetry of the lattice which does not change on cooling below $T_{\rm N}$, one expects the coexistence of degenerate orthogonal AF domains where one type of domain has the characteristics just described, and the second type has the Gd magnetic moments aligned along the $a$-axis with an AF propagation vector ${\bf k}_2=(0, \frac{1}{2},\frac{1}{2})$~r.l.u.  The existence of these two domains has a strong influence on the measured parallel  susceptibility $\chi_{b}(T)$ for $T< T_{\rm N}$.

Anisotropic $\chi(T)$ data~\cite{Dung2009} for a ${\rm GdCu_2Si_2}$ single crystal are shown in Fig.~\ref{Fig:Dung_GdCu2Si2_chi}.  The high-$T$ data follow the Curie-Weiss law with the effective moment and Weiss temperature listed in Fig.~\ref{Fig:Dung_GdCu2Si2_chi}(a).  The effective moment of $7.76\,\mu_{\rm B}$/Gd agrees within about 2\% with the value $\mu_{\rm eff} = g\sqrt{S(S+1)}\mu_{\rm B} = 7.94\,\mu_{\rm B}$ expected for $S = 7/2$ and $g=2$.  An expanded plot of the $\chi_a$ and $\chi_c$ data below 30~K is shown in Fig.~\ref{Fig:Dung_GdCu2Si2_chi}(b).  AF ordering is clearly seen in the $\chi_a(T)$ data at $T_{\rm N} = 12.1$~K, whereas $\chi_c$ is nearly independent of $T$ below $T_{\rm N}$.  The nearly $T$-independent behavior of $\chi_c$ indicates that the ordered magnetic moments are aligned perpendicular to the $c$~axis and therefore lie within the $ab$~plane.  One observes that $\chi_a(T\to0)/\chi_a(T_{\rm N}) \approx 1/2$.  Since the AF structure is known to be collinear within the $ab$~plane as discussed above, this behavior indicates the presence of AF domains as described in the previous paragraph.  

Therefore we fitted the $\chi_a(T<T_{\rm N})$ data by the average of perpendicular and parallel susceptibilities for collinear antiferromagnets, i.e., 
\bea
\chi_{a}(T) &=& \frac{1}{2}[\chi_{\perp} + \chi_\parallel(T)]\nonumber\\*
&=&\frac{1}{2}\left[1+\frac{\chi_\parallel(T)}{\chi(T_{\rm N})}\right]\chi(T_{\rm N}),
\label{Eq:GdCu2Si2Fit}
\eea
where we used Eq.~(\ref{Eq:chiparFit}) for $\chi_\parallel(T)/\chi(T_{\rm N})$ together with the parameters $S = 7/2$ and $f=-1.78$ listed in Fig.~\ref{Fig:Dung_GdCu2Si2_chi}(b).  The fit to the $\chi_a(T<T_{\rm N})$ data is shown in Fig.~\ref{Fig:Dung_GdCu2Si2_chi}(b) and is seen to be reasonably good.  The positive deviation of the fit from the data is typical of such fits by MFT\@.

\subsection{\label{Sec:CuO} CuO: A Quasi-One-Dimensional Spin-1/2 Collinear Antiferromagnet}

\begin{figure}
\includegraphics [width=3.in]{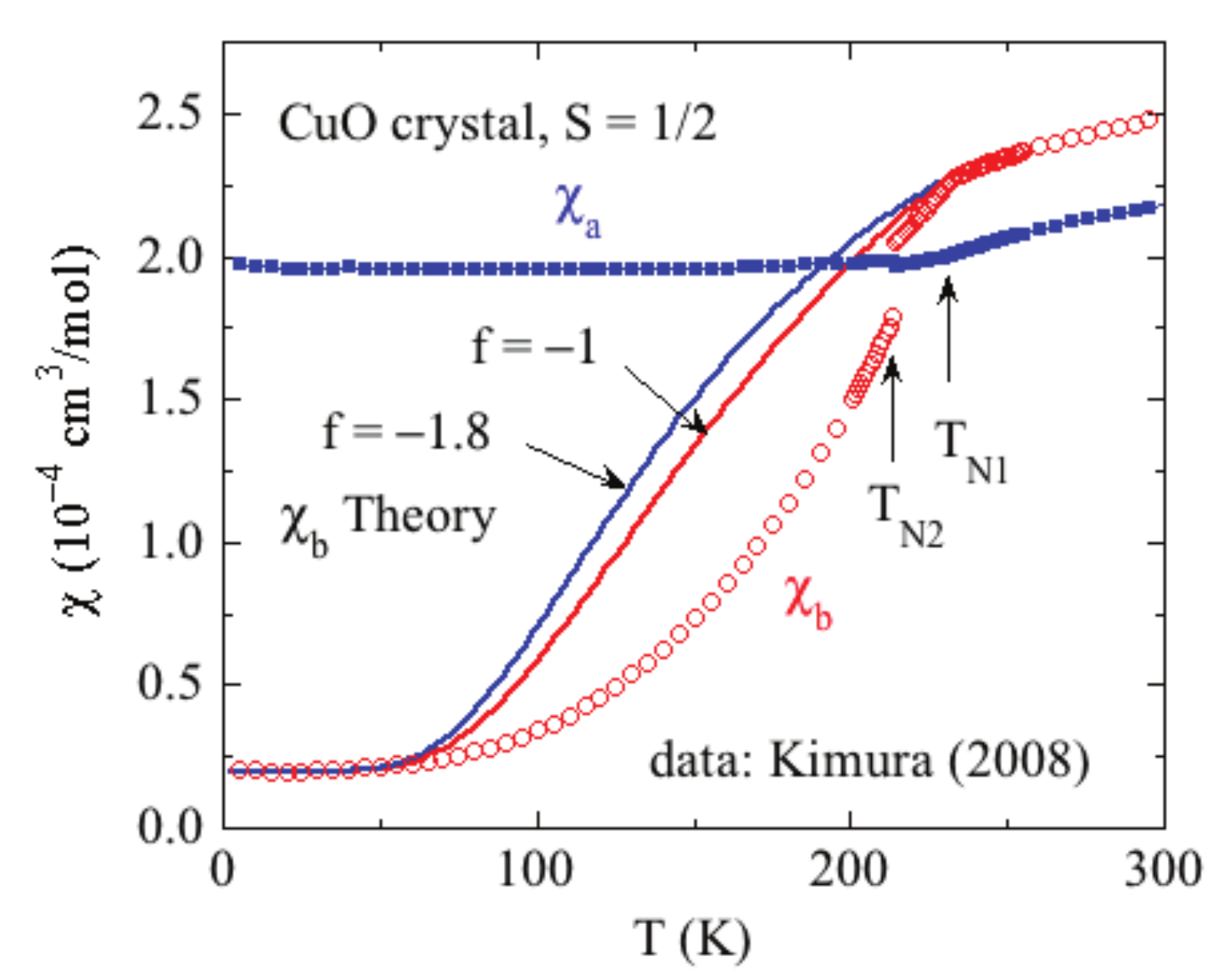}
\caption{ Experimental anisotropic magnetic susceptibilities $\chi_a$ and $\chi_b$ versus temperature $T$ for monoclinic CuO single crystals (symbols)~\cite{Kimura2008}.  A MFT prediction of the easy-axis \mbox{($b$-axis)} $\chi_b(T)$ obtained using Eq.~(\ref{Eq:chiparFit}) with spin $S = 1/2$ and $f = \theta_{\rm p}/T_{\rm N1} = -3.7$ is shown by the solid red curve.  For the prediction, we assumed that the offset of $\chi_b$ from zero at low temperatures is due to the net orbital susceptibility where the spin susceptibility is zero.  The perpendicular susceptibility $\chi_a$ is predicted to be nearly independent of $T$ below $T_{\rm N1}$, as observed.  The anisotropy between $\chi_a$ and $\chi_b$ for $T\geq T_{\rm N1}$ is due to the combined effects of an anisotropic orbital Van Vleck contribution to the susceptibility and an anisotropic $g$-factor.}
\label{Fig:Kimura_CuO_Chi_FitB}
\end{figure}

The insulating compound CuO has a monoclinic structure containing Cu$^{+2}$ spins $S = 1/2$.  The structure consists of Cu-O chains running along the $b$~axis.  Below a second-order magnetic transition at $T_{\rm N1}=230$~K an incommensurate noncollinear AF structure is observed, and below a first-order transition at $T_{\rm N2}=213$~K a collinear AF structure occurs with the ordered moments oriented along the $b$~axis~\cite{Kimura2008}.  The anisotropic susceptibility data~\cite{Kimura2008} are plotted in Fig.~\ref{Fig:Kimura_CuO_Chi_FitB}.  At higher temperatures, a broad maximum in $\chi$ occurs at about 540~K~\cite{Okeeffe1962} that reflects the onset of strong dynamic short-range AF correlations in a low-dimensional spin lattice on cooling.

 $^{63}$Cu NMR shift measurements versus temperature were carried out for a magnetically-aligned powder sample~\cite{Shimizu2003}, from which the authors deduced the values of the anisotropic $g$-factors and Van Vleck orbital susceptibilities.  Then they analyzed the high-$T$ $\chi(T)$ data~\cite{Okeeffe1962} using 1D and 2D spin lattice models and concluded that the data were consistent with a 1D model with an intrachain exchange interaction of 850~K, which is also the Weiss temperature $-\theta_{\rm p}$ for $S=1/2$.  Thus one obtains $f = \theta_{\rm p}/T_{\rm N1}=-3.7$ assuming that interchain interactions are much smaller than intrachain interactions.

Using this value of $f$ and $S=1/2$, the predicted $\chi_b(T\leq T_{\rm N1}$) was computed using Eq.~(\ref{Eq:chiparFit}) and the result is shown in Fig.~\ref{Fig:Kimura_CuO_Chi_FitB}.  A large discrepancy between the observed $\chi_b(T\leq T_{\rm N1}$) and that predicted by MFT is seen.  On the other hand, the perpendicular susceptibility $\chi_a$ is independent of $T$ below $T_{\rm N2}$, in agreement with MFT\@.

Thus with decreasing spin from $S = 7/2$ in Fig.~\ref{Fig:Dung_GdCu2Si2_chi}(b) to $S = 1/2$ in Fig.~\ref{Fig:Kimura_CuO_Chi_FitB}, the experimental $\chi_\parallel(T)$ data increasingly deviate from the MFT predictions.  This suggests an increasing influence of quantum fluctuations on $\chi_\parallel(T)$ with decreasing $S$ as expected.  The increase in quantum fluctuations with decreasing spin is particularly noticable in $\chi_\parallel(T)$ for CuO because the quasi-one-dimensionality of the spin lattice is an additional source of such fluctuations not treated by MFT\@.  It remains unexplained, however, why the $\chi_\parallel(T\leq T_{\rm N1}) = \chi(T_{\rm N1})$ MFT prediction is larger than the experimental data in the temperature range 75--200~K\@.  Further investigation of this issue is warranted.

\subsection{\label{Sec:NoncollinearFits}  120$^\circ$ Coplanar Ordering in Triangular-Lattice and Helical Antiferromagnets}

\begin{figure}
\includegraphics [width=1.75in]{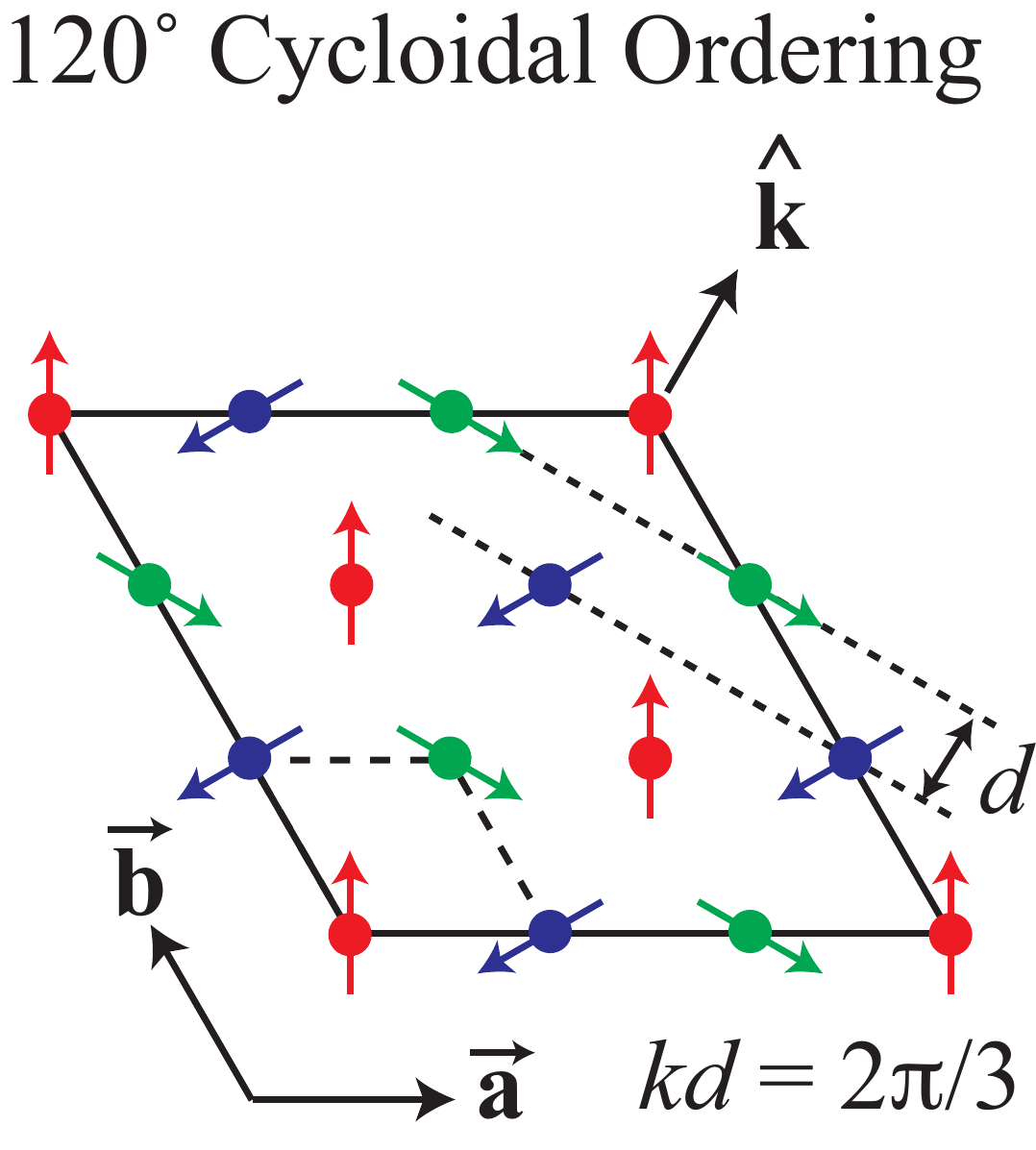}
\caption{ Classical $120^\circ$ ordering on the coplanar simple-hexagonal spin lattice (triangular lattice) illustrated for a cycloidal AF structure with a commensurate wavelength of 3$a$/2~\cite{Johnston2016}.  The hexagonal lattice parameters {\bf a} and {\bf b} ($a=b$) and the direction $\hat{\bf k}$ of the cycloid wave vector {\bf k} are indicated.  The long-dashed line is the outline of the hexagonal unit cell containing one spin and the solid line is the outline of the magnetic unit cell containing nine spins (nine unit cells).  The quantity $d$ is the distance between lines of ferromagnetically-aligned magnetic moments along the cycloid axis ($\hat{\bf k}$) direction and {\bf k} is the AF propagation vector.  The rotation angle of the magnetic moments between adjacent lattice lines in the $\hat{\bf k}$ direction is $\phi_{ji} = kd =2\pi/3$~rad.}
\label{Fig:Triangular_Lattice_AF_k}
\end{figure}

Coplanar helical AF ordering with a turn angle of 120$^{\circ}$ is generically illustrated in Fig.~1 of~\cite{Johnston2012} and cycloidal 120$^{\circ}$ AF ordering in triangular-lattice AFs is depicted in Fig.~\ref{Fig:Triangular_Lattice_AF_k}~\cite{Johnston2016}.   As noted above, the MFT predictions for $\chi(T)$ are identical for the two types of structures.   The remarkable prediction is that $\chi(T\leq T_{\rm N})$ is isotropic, independent of $T$ and also independent of the spin $S$\@.  This prediction was previously verified for the triangular $S=2$ antiferromagnet YMnO$_3$ and the $S=1/2$ antiferromagnet RbCuCl$_3$~\cite{Johnston2012}.  Here two additional such compounds are considered.

\begin{figure}
\includegraphics [width=3.in]{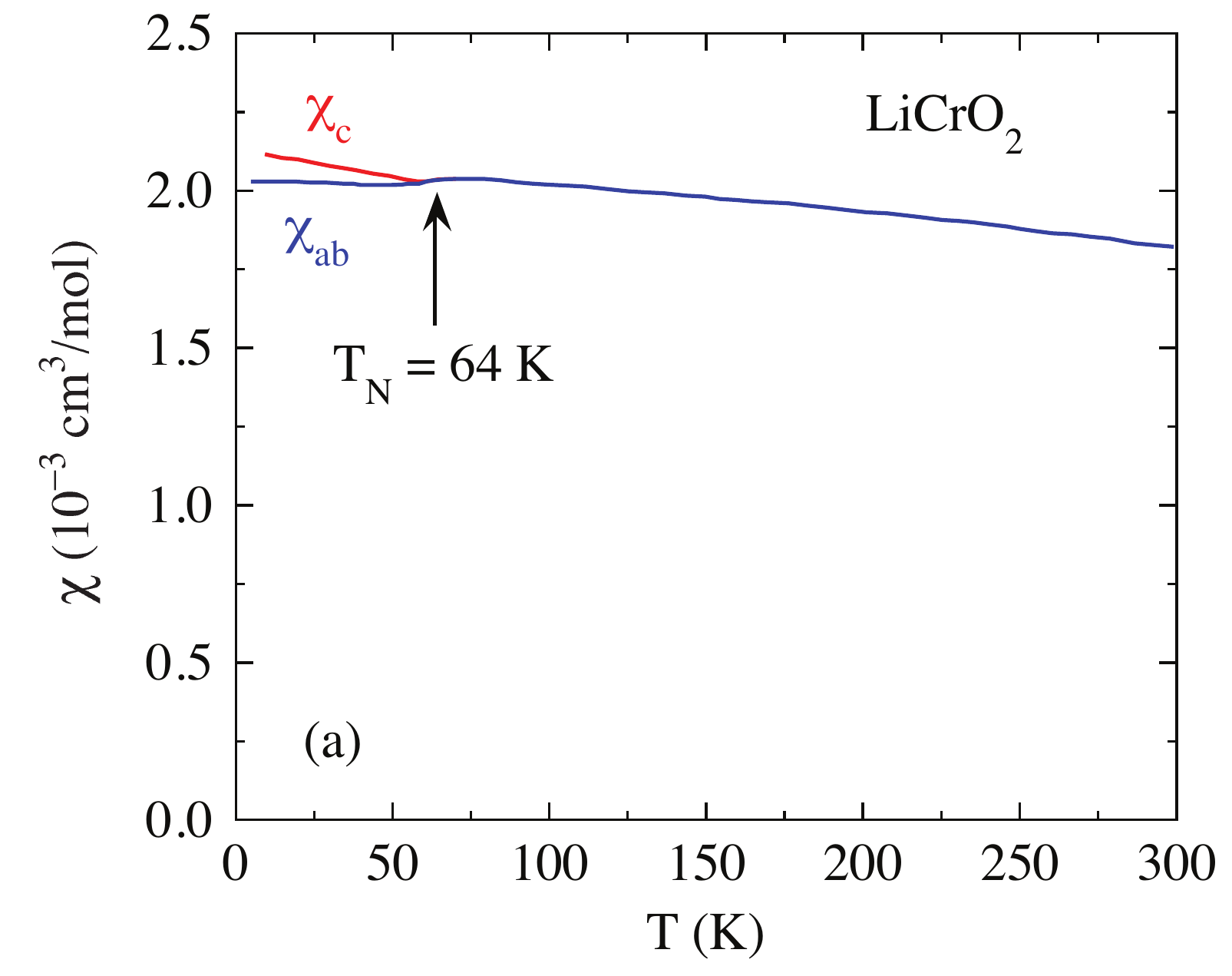}
\includegraphics [width=3.in]{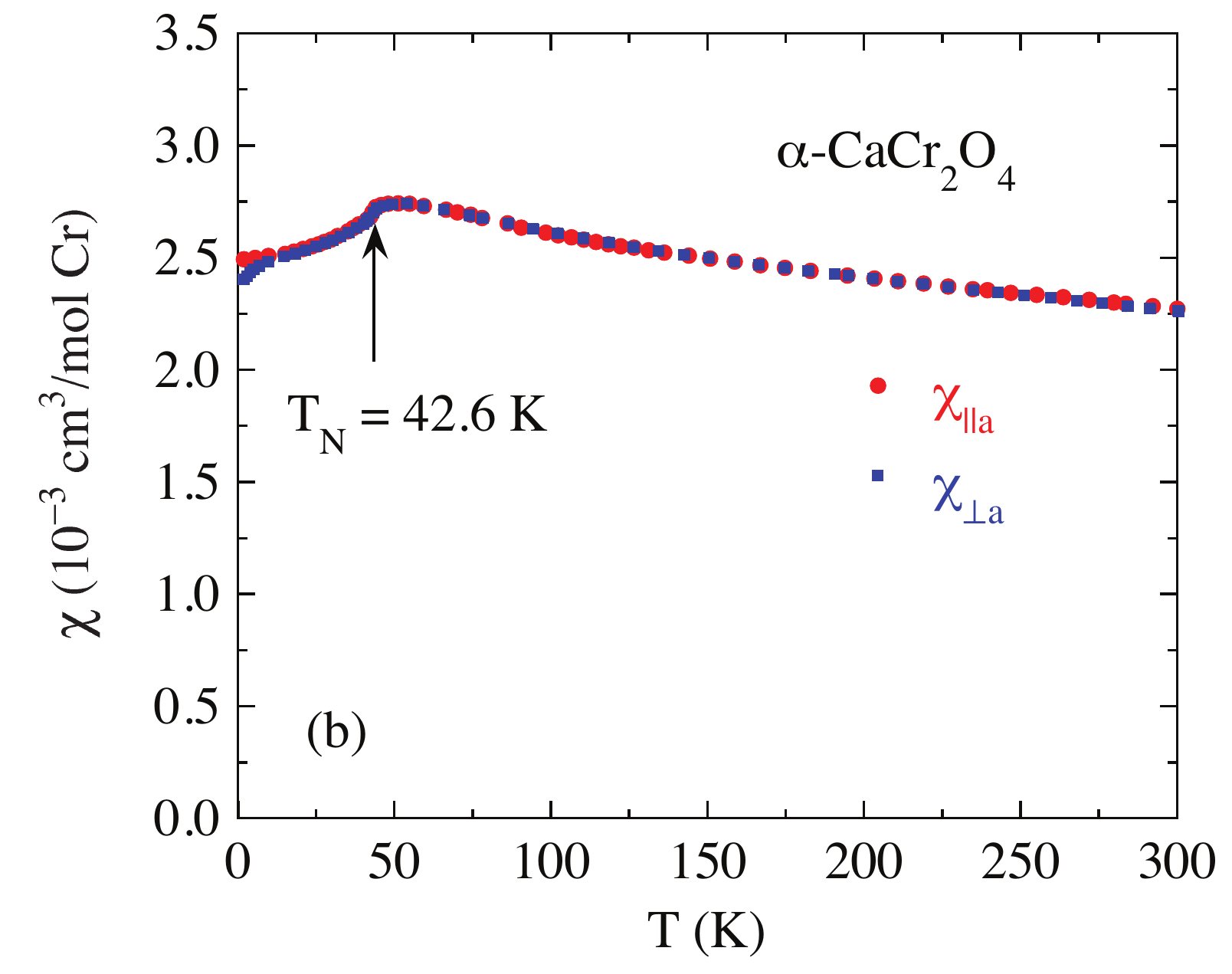}
\caption{ Experimental anisotropic susceptibilities $\chi_{ab}$ and $\chi_c$ versus temperature $T$ for triangular antiferromagnets (a)~${\rm LiCrO_2}$ with spin $S = 3/2$~\cite{Kadowaki1995} and (b)~$\alpha$-${\rm CaCr_2O_4}$ with $S = 3/2$~\cite{Toth2011}.  It was not possible to obtain data along the helix axis in (b) (i.e., $\perp a$), because of three-fold twinning about the $a$-axis~\cite{Toth2011}.  The plotted data were digitized from the published figures.  The nearly isotropic and temperature-independent spin susceptibility below $T_{\rm N}$ predicted within MFT is a signature of  120$^\circ$ coplanar cycloidal spin ordering such as in Fig.~\ref{Fig:Triangular_Lattice_AF_k}. }
\label{Fig:LiCrO2_YMnO3_chi}
\end{figure}

Anisotropic $\chi(T)$ data are shown in Fig.~\ref{Fig:LiCrO2_YMnO3_chi} for a single crystal of the $S=3/2$ triangular-lattice antiferromagnet LiCrO$_2$ with $120^\circ$ coplanar cycloidal ordering~\cite{Kadowaki1995} and for a single crystal of \mbox{$\alpha$-${\rm CaCr_2O_4}$} that exhibits coplanar 120$^\circ$ helical ordering~\cite{Toth2011}.  The authors of Ref.~\cite{Kadowaki1995} provided no fit to the $\chi(T)$ data for LiCrO$_2$ above $T_{\rm N} = 64$~K\@.  The authors of Ref.~\cite{Toth2011} fitted the isotropic $\chi(T)$ data for $\alpha$-${\rm CaCr_2O_4}$ from 800 to 1000 K by the modified Curie-Weiss law, far above $T_{\rm N} = 42.6$~K, yielding the paramagnetic Weiss temperature $\theta_{\rm p} = -564(4)$~K and effective magnetic moment $\mu_{\rm eff} = 3.68(1)\,\mu_{\rm B}$/Cr, close to the spin-only value of $3.87\,\mu_{\rm B}$/Cr obtained assuming spin $S = 3/2$ and a spectroscopic splitting factor $g = 2$.

The compound $\alpha$-${\rm CaCr_2O_4}$ has a slight orthorhombic distortion from a triangular-lattice structure~\cite{Toth2011}.  The helix axis is directed along the orthorhombic $b$-axis direction with the magnetic moments in the $ac$-plane and with a magnetic moment turn angle between adjacent planes of the helix of $kd = 119.86(2)^\circ$~\cite{Chapon2011}.  Note that the AF propagation vector~{\bf q} quoted by the magnetic x-ray and neutron scattering community is not necessarily the same as the wave vector~{\bf k} of the helix or cycloid.  In Ref.~\cite{Chapon2011}, {\bf q} is given as $0.3317(2)~(2\pi/b)\hat{\bf b}$.  Using $d=b/4$ gives the turn angle $qd\approx 30^\circ$ instead of $120^\circ$.  What happened is that ${\bf k} =1.3317(2)(2\pi/b)\hat{\bf b}$ was changed to $0.3317(2)(2\pi/b)\hat{\bf b}$ in order to translate {\bf q} by the reciprocal lattice translation vector $-(2\pi/b)\hat{\bf b}$ into the first Brillouin zone that extends along the $b$-axis direction from $-\pi/b$ to $+\pi/b$.

Each of the compounds LiCrO$_2$ and $\alpha$-${\rm CaCr_2O_4}$ shows nearly isotropic and $T$-independent $\chi(T\leq T_{\rm N})$ behavior as seen in Figs.~\ref{Fig:LiCrO2_YMnO3_chi}(a) and~\ref{Fig:LiCrO2_YMnO3_chi}(b), respectively~\cite{Kadowaki1995, Toth2011}.  Similar $\chi(T\leq T_{\rm N})$ behavior has also been observed for many other triangular lattice antiferromagnets with $120^\circ$ cycloidal ordering, such as the $S = 3/2$ systems VF$_2$ and VBr$_2$~\cite{Hirakawa1983}.  These experimental results confirm the MFT prediction that $\chi(T\leq T_{\rm N})$ for antiferromagnets showing 120$^\circ$ coplanar helical or cycloidal ordering is (approximately) isotropic and independent of $f$, $S$ and~$T$\@.

\subsection{\rm\bf Heat Capacity of a Collinear Antiferromagnet with Large Spin: GdNiGe$_3$}

The compound GdNiGe$_3$ crystallizes in an orthorhombic structure with space group \emph{Cmmm} with lattice parameters $a = 4.0551(2),\ b=21.560(2),\ c= 4.0786(7)$~\AA~\cite{Ko2008, Mun2010}.  The Gd sublattice consists of slightly orthorhombically distorted square lattices in the $ac$~plane that are stacked along the $b$~axis.  

The anisotropic $\chi(T)$ data for a single crystal of this compound were presented in Ref.~\cite{Mun2010} and were fitted rather well by MFT in Ref.~\cite{Johnston2012}.  The former authors determined from a Curie-Weiss fit to the high-$T$ susceptibility data at $T>T_{\rm N}$ that the magnetism in this compound arises from Gd spins $S=7/2$ with \mbox{$g=2$}.  The Weiss temperature in the Curie-Weiss law is \mbox{$\theta_{\rm p} = -23.0$~K} and the Gd spins order antiferromagnetically at $T_{\rm N} = 26.2$~K, yielding $f = \theta_{\rm p}/T_{\rm N} = -0.88$.  It is clear from the data for $T < T_{\rm N}$ that GdNiGe$_3$ is a collinear antiferromagnet with the $a$-axis being the easy axis, but the detailed magnetic structure has not been determined.  Irrespective of that,  Eq.~(\ref{Eq:chiparFit}) still applies for fitting $\chi_\parallel(T<T_{\rm N})$, which illustrates the utility of the MFT\@.  The perpendicular susceptibility along the $b$~axis in the ordered state with $T \leq T_{\rm N}$ is predicted to be independent of $T$, in good agreement with the $\chi_b(T)$ data.  The fit of the $a$-axis parallel-susceptibility data for GdNiGe$_3$ by MFT using Eq.~(\ref{Eq:chiparFit}) with no adjustable parameters~\cite{Johnston2012} is rather good.  The deviation of the fit from the data is likely due to dynamic magnetic fluctuations and correlations not accounted for by MFT\@.  These correlations are evident in the $\lambda$-shaped peak in $C_{\rm mag}(T)$ at $T_{\rm N}$ with a nonzero contribution above $T_{\rm N}$ as discussed next.

\begin{figure}
\includegraphics [width=2.75in]{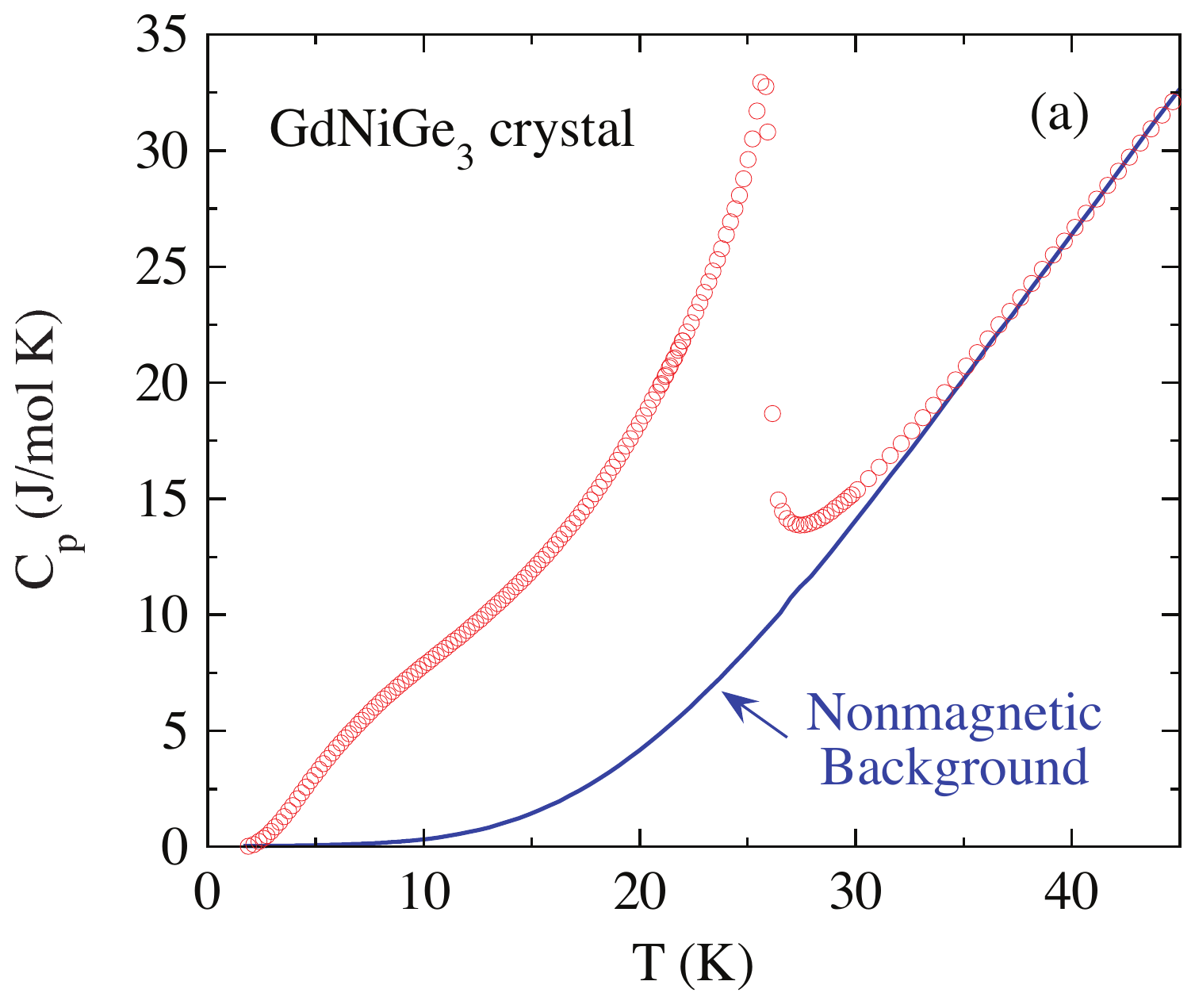}
\includegraphics [width=2.75in]{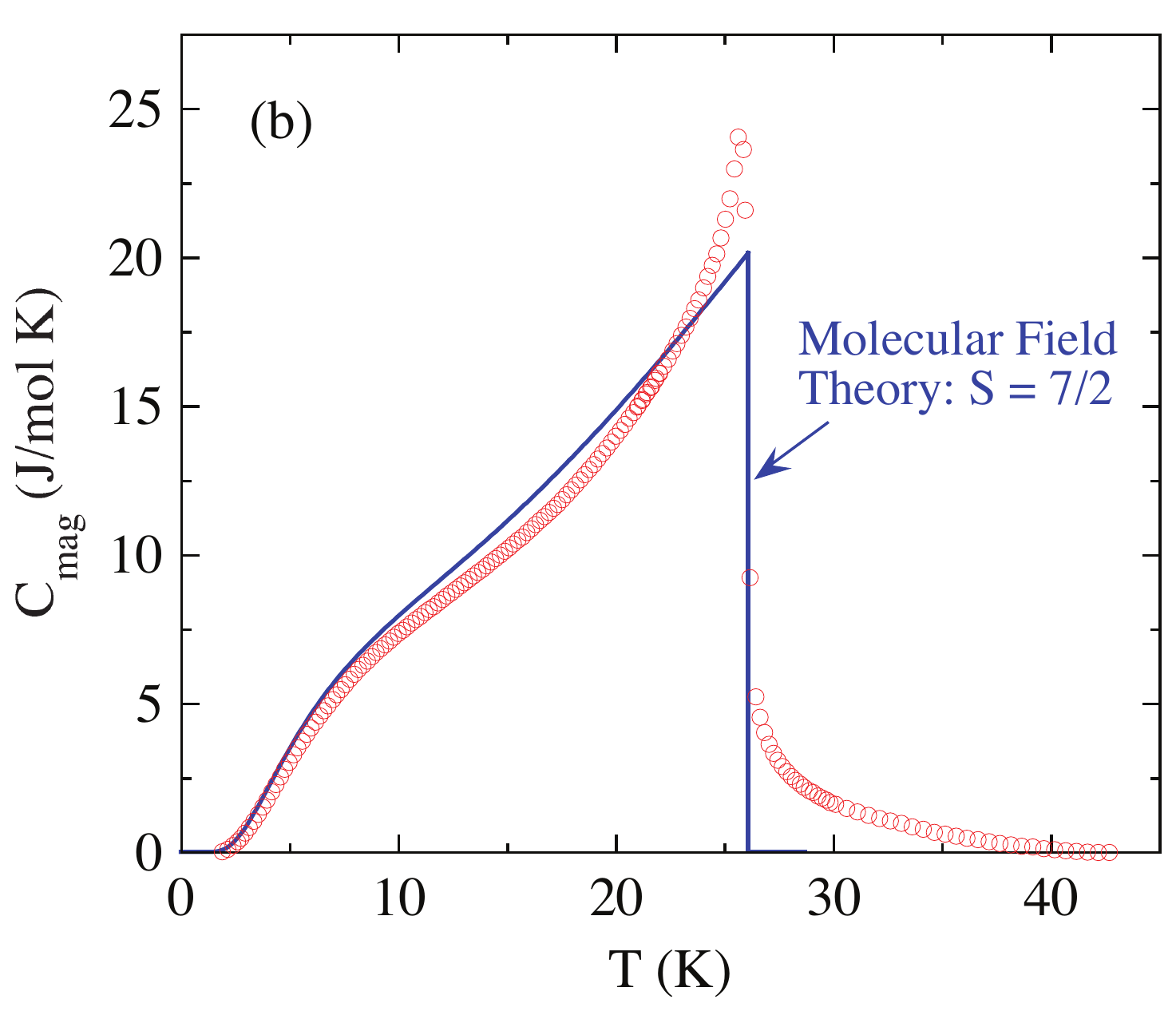}
\caption{ (a) Heat capacity $C_{\rm p}$ versus temperature $T$ of a single crystal of GdNiGe$_3$ in zero magnetic field (open red circles)~\cite{Mun2010}. Also included is the nonmagnetic background we estimated from $C_{\rm p}(T)$ data~\cite{Mun2010} for isostructural YNiGe$_3$ (solid blue curve).  (b) Magnetic contribution $C_{\rm mag}$ to the heat capacity versus $T$ obtained by subtracting the nonmagnetic background contribution from the data in (a).  Also shown in~(b) is the molecular field theory (MFT) prediction of $C_{\rm mag}(T)$ in Eq.~(\ref{Eq:Cmag}) for spin $S = 7/2$ (solid blue curve).  The nonzero $C_{\rm mag}$ for $T > T_{\rm N}$ arises from dynamic short-range AF fluctuations and correlations not taken into account by the MFT theory.}
\label{Fig:GdNiGe3_Cp}
\end{figure}

Due to the large Gd spin $S=7/2$ one expects that the magnetic heat capacity $C_{\rm mag}$ versus $T$ for GdNiGe$_3$ should be nearly mean-field-like.  The zero-magnetic-field heat capacity $C_{\rm p}(T)$ of single-crystal GdNiGe$_3$~\cite{Mun2010} is plotted versus~$T$ in Fig.~\ref{Fig:GdNiGe3_Cp}(a).  The nonmagnetic heat capacity background was estimated here by multiplying the temperatures for $C_{\rm p}(T)$ of the isostructural nonmagnetic reference compound YNiGe$_3$~\cite{Mun2010, Ko2008} by a factor of~0.92.  The  magnetic contribution $C_{\rm mag}(T)$ to $C_{\rm p}(T)$ of GdNiGe$_3$ is then obtained by subtracting  the temperature-normalized $C_{\rm p}(T)$ for YNiGe$_3$ from the measured $C_{\rm p}(T)$ for GdNiGe$_3$ and is plotted in Fig.~\ref{Fig:GdNiGe3_Cp}(b).  The MFT prediction is given by~\cite{Johnston2015}
\bea
\frac{C_{\rm mag}(t)}{R} = \frac{3S\bar{\mu}_0^2}{(S+1)t\Big[\frac{(S+1)t}{3B_S^\prime(y_0)} - 1\Big]},
\label{Eq:Cmag}
\eea
where $R$ is the molar gas constant.  The fit  for spin \mbox{$S = 7/2$} with no adjustable parameters is seen to describe the $C_{\rm mag}(T)$ data in Fig.~\ref{Fig:GdNiGe3_Cp}(b) reasonably well.  The good fits of $\chi_\parallel(T)$ and $C_{\rm mag}(T)$ by MFT are not unexpected since MFT should be fairly accurate for large spins such as for Gd$^{+3}$.  One can still see evidence of dynamic short-range AF correlations both above and below $T_{\rm N}$ (not treated by MFT) by the $\lambda$~shape of $C_{\rm mag}$ below $T_{\rm N}$ and the nonzero $C_{\rm mag}(T)$ above $T_{\rm N}$.

The hump that occurs in $C_{\rm mag}(T)$ in Fig.~\ref{Fig:GdNiGe3_Cp}(b) at {\mbox{$T \sim T_{\rm N}/3$} arises naturally in MFT\@.  The hump becomes more pronounced as the Zeeman degeneracy of the ground state increases~\cite{Johnston2011, Rodriguez2005}, so it is quite pronounced for Gd$^{+3}$ and Eu$^{+2}$.  It arises in MFT from (1) the $T$ dependence of the ordered moment in Eq.~(\ref{Eq:Cmag})  which gives a $T$ dependence to the exchange field seen by each moment, which in turn causes the splitting of the Zeeman levels to depend on $T$, and from (2) the $T$-dependent Boltzmann populations of those levels.  As noted in Ref.~\cite{Johnston2011}, large-$S$ systems must develop the hump in order that the molar magnetic entropy $S_{\rm mag}$ in the disordered state at $T_{\rm N}$ increases with increasing $S$ according to the statistical mechanics requirement $S_{\rm mag} = R \ln(2S+1)$, because for $T \gtrsim T_{\rm N}/3$ and $S\gtrsim 2$ the magnetic heat capacity is limited from above by the classical calculation of $C_{\rm mag}(T)$.  The hump is not as prominent for rare-earth antiferromagnets not containing the $s$-state ions Eu$^{+2}$ or Gd$^{+3}$ with $S=7/2$ because the Hund's-rule ground states of other $R^{+3}$ rare earth ions are split by the crystalline electric fields via the spin-orbit interaction which reduces the Zeeman degeneracies of the zero-field ground states to values too small to cause the hump to form in the magnetically-ordered state.

\section{\label{Sec:Summary} Concluding Remarks}

As discussed in the Introduction, the unified molecular-field theory has been successful in fitting the anisotropic  magnetic susceptibilities $\chi(T)$ below $T_{\rm N}$ and of the magnetic component $C_{\rm mag}(T)$ of the heat capacity for a variety of collinear and coplanar noncollinear magnetic structures in single crystals containing identical crystallographically-equivalent spins interacting by Heisenberg exchange including helical and 120$^\circ$ coplanar structures on triangular lattices.  

We considered the tetragonal compound GdCu$_2$Si$_2$ which exhibits collinear AF ordering with the moments aligned along the $a$ or~$b$ axes.  Due to the tetragonal symmetry, the single-crystal $\chi(T)$ data evidenced the presence of orthogonal AF domains.  Assuming the domain populations were the same, we obtained reasonably good agreement between the observed and calculated in-plane susceptibility $\chi_{ab}(T\leq T_{\rm N})$.

The parallel susceptibility of a single crystal of the monoclinic $S=1/2$ compound CuO with collinear AF order was fitted by the MFT with no adjustable parameters and poor agreement with the experimental data was found.  Although one might attribute this disagreement to the small spin of the Cu$^{+2}$ ion, further work is needed to ascertain the actual origin of this unusually-large discrepancy.  For example, the anisotropic $g$~factor could contribute to it.  The poor fit to the $\chi(T)$ data for CuO sharply contrasts to the typically very good fits obtained using the unified MFT for other antiferromagnets and thus represents a conundrum that would be interesting to investigate further.

Two examples of coplanar AF ordering in Cr$^{+3}$ spin-3/2 compounds were discussed.  In LiCrO$_2$ the Cr spins occupy a coplanar triangular lattice and the AF structure is a $120^\circ$ cycloidal structure, whereas in \mbox{$\alpha$-${\rm CaCr_2O_4}$} the Cr spins order in a helix with a turn angle of 120$^\circ$.  The MFT has the same prediction for the anisotropic $\chi(T)$ of both magnetic structures, namely that {\mbox{$\chi(T\leq T_{\rm N})$} is independent of $T$, and perhaps surprisingly, also of $S$\@.  The susceptibilities of these compounds indeed approximately followed this prediction below their N\'eel temperatues of 64~K and 42.6~K, respectively.

We also extended the range of these fits to include single-crystal $C_{\rm mag}(T)$ data for the collinear antiferromagnet GdNiGe$_3$ with $T_{\rm N}=26.2$~K\@.  This compound shows a $\lambda$-shaped anomaly in $C_{\rm mag}(T)$ at $T_{\rm N}$ which contrasts with the step-like change predicted by MFT\@.  Furthermore, the $C_{\rm mag}(T)$ exhibits a tail above $T_{\rm N}$ arising from dynamic short-range magnetic  ordering of the $S=7/2$ Gd spins not predicted by the MFT\@.  This has also been observed in other $S=7/2$ antiferromagnets such as the helical antiferromagnets EuCo$_2$P$_2$~\cite{Sangeetha2016}, EuCo$_2$As$_2$~\cite{Sangeetha2018}, and EuNi$_2$As$_2$~\cite{Sangeetha2019}.

For cases where reasonably-good agreement of the MFT with the experimental susceptibility data was found, there remain relatively small but systematic deviations of the MFT predictions from the anisotropic $\chi(T)$ data for both collinear and coplanar noncollinear antiferromagnets  containing Heisenberg spins.  It would be interesting and useful to quantitatively establish the origin of these \mbox{deviations}.

\acknowledgments

The author is grateful to S.~L.~Bud'ko and T.~Kimura for communicating experimental data.  This work was supported by the U.S. Department of Energy, Office of Basic Energy Sciences, Division of Materials Sciences and Engineering.  Ames Laboratory is operated for the U.S. Department of Energy by Iowa State University under Contract No.~DE-AC02-07CH11358.\\


\end{document}